\documentclass[11pt]{article}
\usepackage{graphics}
\usepackage{graphicx,epsf,rotate}   % needed for figures
\usepackage{epsfig}
\usepackage{tabularx}
\begin{document}

\title{A soft ellipsoid potential for biaxial molecules : Modeling
and computer simulation}

\author{Jayashree Saha  \\
Department of Physics, University of Calcutta\\
92, A.P.C. Road, Kolkata - 700009, India\\
E-mail : jsphy@caluniv.ac.in}

\maketitle

\begin{abstract}

A soft ellipsoid contact potential model for a pair of 
biaxial ellipsoidal molecules is proposed  which  considers 
the configuration dependent energy anisotropy explicitly along with
their geometrical aspects. We performed Molecular Dynamics simulation 
study to generate both biaxial smectic and nematic phases using  
this new potential.

\end{abstract}

%`\pacs{Valid PACS appear here}% PACS, the Physics and Astronomy

\newpage
%%%%%%%%%%%%%%%%%%%%%%%%%%%%%%%%%%%%%%%%%%%%%%%%%%%%%%%%%%%%%%%%%%%%%%%%%%%
%%%%%%%                       BODY OF TEXT
%%%%%%%%%%%%%%%%%%%%%%%%%%%%%%%%%%%%%%%%%%%%%%%%%%%%%%%%%%%%%%%%%%%%%%%%%%%

%\section{Introduction}
%\label{sec:intro}

Liquid crystal forming compounds are complex organic molecules, having   in general, cylindrically symmetric rod-like or disc-like structures,. Theoretical models, developed to study phase transitions, structures and dynamics of these systems, essentially strive to relate important molecular interactions to the experimentally observed phase behaviour and properties. In this respect, computer simulation is one of the most efficient
methods which considers detailed microscopic interactions theoretically to  explain and analyze macroscopic experimental observations. Non-polar liquid crystal molecules interact both through short-range hard interactions and long range attractive interactions originated primarily 
from electrostatic interactions \cite{Luck_book}. To study a liquid crystal system using computer simulation two major steps are to be followed. The first one is the modelling of a suitable potential which takes into account the basic electrostatic interactions responsible for the generation of one (or more) specific phase(s) and the second step is to use an appropriate simulation method. These models may be broadly classified as full atomic, united atom, site-site and single site interaction models. The first three class of models are not computationally efficient as there
are multi-site interaction terms present in the potential. The fourth type of model potentials can be used efficiently for computer simulation studies of these complex molecular systems in a realistic way. 

Single-site potentials require determination of $\sigma$, distance of closest approach as a function of orientation and center of mass distance for a pair of molecules and it is a difficult task to estimate this correctly in case of  anisotropic molecular systems like liquid crystals. Additionally, the potentials must capture both the geometric and energy anisotropies of the systems consistently. To develop single-site model potentials which are accurate, relatively simple and computationally efficient, a rod-like rigid molecule may be approximated as a prolate ellipsoid. The main idea behind it is the representation of the interactions between a pair of liquid crystal molecules by their joint probability, where each molecule has Gaussian charge distribution centred around the molecular centroids. Corner \cite{Corner}developed a model potential based on this idea which makes the analytical calculations remarkably simplified. Further significant development of this class of potential was done by Berne's group  which led to widely used
Gay-Berne(G-B) potential \cite{Gay81}. These models were basically modified form of Lennard-Jones potential where the energy strength parameter $\epsilon$ and energy range parameter $\sigma$  depended on molecular orientations and positions. In this form it provided both attractive and repulsive contributions of the intermolecular interactions of rigid, uniaxial molecules, in a computationally tractable way. These potentials known as Gaussian overlap potentials (GOP) have become popular in simulation studies due to their comparatively simplified approach in determining the energy contribution of a pair of ellipsoidal molecules. 
However, the G-B potential in its original form being generally restricted to uniaxial molecules is not an effective model when the multiplicity of phases adopted by liquid crystal molecules are to be studied. Modifications \cite{Cleaver96} and some parametrizations \cite{zan_rev} of Gay-Berne potential have been proposed in order to study the behaviour of biaxial phases but these GOP models are not in conformity with the actual geometry of the molecule. The range function and the energy strength function in these potentials do not correctly estimate the interactions between the non-uniaxial ellipsoidal particles. Most of the 
real liquid crystal molecules do not have cylindrical uniaxial structures, 
therefore, biaxial contribution to the potential is essential when one
studies various phase structures adopted by the real molecules. Though much progress in the experimental study have been achieved, there has been comparatively little theoretical work done on this important class of liquid crystal forming molecules which have biaxial structures. 

Perram-Wertheim \cite{Perram84} ellipsoid contact potential (ECP) had advantages over Gaussian overlap type potential because the ECP calculated geometrical contact point accurately for ellipsoidal structures including biaxial molecules. ECP represented the ellipsoids as quadratic forms to solve the contact function accurately by doing
an optimization calculation. Despite this fact, ECP has not yet been widely 
used to study liquid crystals because it did not differentiate in energies 
due to different molecular configurations, thereby 
neglecting the proper weightage of the energy strength contribution to the potential. Energy strength parameter $\epsilon$ was taken as a constant in the ECP therefore well depth in this attractive-repulsive potential did not show correct behaviour. In our earlier work \cite{saha11}, we modelled a soft ellipsoid contact potential for uniaxial
ellipsoidal molecules, considering the configuration dependent energy anisotropy along
with their geometrical aspects. In this paper we generalize the earlier model by including biaxial contribution in the pair interaction potential for ellipsoidal molecules. 
Explicit analytical expressions for proper range and strength 
parameters alongwith related forces and torques are also provided. In addition, the molecular dynamics simulation study shows that this new soft ellipsoidal pair potential for biaxial molecules is suitable for generating both biaxial
and uniaxial liquid crystal phases successfully in a single component rod-like molecular system. 
  
%\section{Model}
%\label{sec:model}

In our model the charge distribution of molecules are described as ellipsoidal and are approximately represented by 3-D gaussian functions, i.e. the density of the centre of force along the axis of a molecule is a Gaussian function of distance from the center of the ellipsoid. The total interaction of two molecules is again a Gaussian distribution and is a function of their intercenter distance and relative orientations. 

The pair interaction potential of two ellipsoidal molecules can be evaluated
from the joint probability distribution of two Gaussians following convolution
integral principle. This joint distribution function is proportional to
\[
\exp [{\bf r^\mathsf{T}} . {\bf H}^{-1} . {\bf r}] 
\]
where $\bf r$ is the intermolecular vector connecting the centers of the ellipsoids and the matrix ${\bf H}$ depends on the axes lengths and the relative orientations of three principal semi-axes of the two interacting ellipsoids.
Both the energy range function $\sigma$ i.e. the permitted closest distance between molecular pairs and the energy strength function $\epsilon$ which gives estimation of potential depths and widths, depend on this form of joint probability distribution functions.

The model pair potential which is a modified version of Lennard-Jones type interaction potential for the prolate ellipsoidal molecules can be represented in a shifted form as

\newpage

\begin{eqnarray}
U = 4 \epsilon( \hat{\alpha}, \hat{\beta}, {\hat r}) [ (\frac{\sigma_{0}}{r-\sigma( \hat{\alpha}, \hat{\beta}, {\hat r})+\sigma_{0}})^{12} -\nonumber \\
(\frac{\sigma_{0}}{r-\sigma( \hat{\alpha}, \hat{\beta}, {\hat r})+\sigma_{0}})^{6}].
\end {eqnarray}

where, $\hat{\alpha}$ and $\hat{\beta}$, are the orthogonal rotation matrices for transformation from space-fixed to molecule fixed coordinates for molecules 1 and 2 respectively and the value of $\sigma_{0}$ is equal to the smallest semi-axis length of the ellipsoid. 

Unlike constant $\epsilon$ parameter in ECP model, our model considers   relative configuration dependent $\epsilon$ for a
molecule pair because they are characterzised by highly non-spherical charge distributions. 
The energy strength function $\epsilon$ in our model potential is taken as
\begin{equation}
\epsilon = \epsilon_{0} \epsilon_{1}^{\nu}(\hat{\alpha}, \hat{\beta}) 
\epsilon_{2}^{\mu}(\hat{\alpha}, \hat{\beta}, {\hat r}).  
\end{equation}
Here, $\epsilon_{0}$ is a constant term and $\mu$, $\nu$
are adjusting parameters for relative well depth variation in different 
compounds; $\epsilon_{1}$ is a function of 
$\hat{\alpha}$ and $\hat{\beta}$; $\epsilon_{2}$ 
depends additionally on $\hat{r}$, which is the unit vector along the 
intermolecular vector connecting the centers of the ellipsoids.\\

To calculate $\epsilon_{2}$, we consider diagonal energy matrix for a biaxial molecule as 
\begin{equation}
{\gamma} = 
\left(
\begin{array}{ccc}
(\frac{\epsilon_{0}}{\epsilon_{x}})^{\frac{1}{\mu}} & 0 & 0 \\
0& (\frac{\epsilon_{0}}{\epsilon_{y}})^{\frac{1}{\mu}} & 0\\
0 & 0 & (\frac{\epsilon_{0}}{\epsilon_{z}})^{\frac{1}{\mu}} \\
\end{array}
\right).
\end{equation}
where $\epsilon_{x}, \epsilon_{y}, \epsilon_{z}$ are equal to the 
well depths for side by side (s-s), width to width (w-w) and end to end (e-e) configurations and for the biaxial model we take $\epsilon_{x}>\epsilon_{y}>\epsilon_{z}$. We also take $\epsilon_{0} = \epsilon_{x}$.

Defining matrices $\bf A$ and $\bf B$ which correspond to the first and second molecule respectively as
\begin{equation}
%\begin{array}{rcl}
{\bf A} =  \hat{\alpha}^\mathsf{T} { \gamma}\hskip 0.1 cm   \hat{\alpha} ;
\hskip 0.6 cm 
{\bf B} =  \hat{\beta }^\mathsf{T} {\gamma}\hskip 0.1 cm   \hat{\beta }. 
\end{equation}

We write 
\begin{equation}
{\bf H_{E}} = \lambda_{E} {\bf A} + (1-\lambda_{E}) {\bf B}.
\end{equation}
where $\lambda_{E}$ is an energy related adjustable parameter such that 
$\lambda_{E} \in [0,1]$ and this  makes the 
object function
 \begin{equation}
S_{E}(\lambda_{E}) = \lambda_{E}(1-\lambda_{E}) \hat{r}^\mathsf{T} \cdot {\bf H_{E}}^{
-1} \cdot 
\hat{r}.
\end{equation}
an optimum. This object function gives correct estimation of the energy strength
function by properly incorporating asymmetric configurations of a biaxial molecule pair.

The $\hat r$ dependent part of the energy strength function $\epsilon_{2}$ in our model potential is taken as

\begin{eqnarray}
\epsilon_{2}(\hat{\alpha}, \hat{\beta}, \hat{r}) =
\lambda_{E}(1-\lambda_{E}) \hat{r}^\mathsf{T} \cdot [(\lambda_{E} {\bf A} + 
(1-\lambda_{E}) {\bf B}]^{-1} \cdot \hat{r} \nonumber \\
= \lambda_{E}(1-\lambda_{E}) \hat{r}^\mathsf{T}\cdot {\bf H_{E}}^{-1} \cdot \hat{r} \nonumber \\.
\end{eqnarray}
  
We take $\epsilon_{1}$ for biaxial molecules 
as \cite{Berardi95}                       

\begin{equation}
\epsilon_{1}(\hat{\alpha}, \hat{\beta}) =  \sigma_{t}  |\bf{H_{E}^{'}}|^{-1/2}.
\end{equation}
where $\bf{H_{E}^{'}} = \frac{1}{2} (\bf{A}+ \bf{B})$ and 
$\sigma_{t} = {\sqrt{2\frac{\sigma_{y}}{\sigma_{x}}}}\sigma_{0}^{3}[(\frac{\sigma_{y}}{\sigma_{x}})+(\frac{\sigma_{z}}{\sigma_{x}})^{2}]$;
$\sigma_{x}$, $\sigma_{y}$ and $\sigma_{z}$ are the values of the range function for s-s, w-w and e-e configurations respectively. We calculate $\sigma$ following ECP \cite{Perram84}. 

Defining matrices ${\bf M}$ and ${\bf N}$  related to the first and the second molecule respectively as   
\begin{equation}
%\begin{array}{rcl}
{\bf M} =  \hat{\alpha}^\mathsf{T} {\bf S}^{2} \hskip 0.1 cm   \hat{\alpha} ;
\hskip 0.6 cm 
{\bf N} =  \hat{\beta }^\mathsf{T} {\bf S}^{2} \hskip 0.1 cm   \hat{\beta } 
\end{equation}
where diagonal shape matrix  ${\bf S}$ takes the folowing form
\begin{equation}
{\bf S} = 
\left(
\begin{array}{ccc}
\sigma_{x} & 0 & 0 \\
0& \sigma_{y}& 0\\
0 & 0 & \sigma_{z} \\
\end{array}
\right).
\end{equation}
and  
${\bf H_{D}} = \lambda_{D}{\bf M}+(1-\lambda_{D}){\bf N}$.
We express the distance of closest approach for biaxial molecules
$\bf \sigma$ in the following form 
\begin{equation}
\sigma^{-2} = \lambda_{D}(1-\lambda_{D}) ({\hat r^{T}} . {\bf H_{D}}^{-1} . {\hat r}). 
\end{equation}
where $\lambda_{D}$ is the shape related adjustable parameter.
For our biaxial model $\sigma_{x} < \sigma_{y} < \sigma_{z}$.
The potential is not isotropic at large distance which
gives small thermodynamic contribution and therefore neglected for simplicity. The forces and torques are derived analytically for the biaxial ellipsoid. The 
expressions for force ${\vec f}$ and torque ${\vec \tau}$ are
\begin{eqnarray}
{\vec f}  =  \frac{8 \epsilon_{0} \epsilon_{1}^{\nu} \epsilon_{2}^{\mu}}
{\sigma_{0}
r^{2}} [(3 r^{2}(2 \rho^{-13}
-\rho^{-7}){\hat r} + 3 \sigma^{3} \lambda_{D}(1-\lambda_{D})\nonumber \\
 (2\rho^{-13}
-\rho^{-7}) \{  {\bf k} - (\hat r .   {\bf k}) \hat r \}  
- \lambda_{E}(1-\lambda_{E}) \nonumber \\
\mu \epsilon_{2}^{-1} \sigma_{0}  (\rho^{
-12} -\rho^{-6})
 \{{\bf k}_{E} - (\hat r .  {\bf k}_{E}) \hat r\}] \nonumber \\.
\end{eqnarray}

\begin{eqnarray}
{\vec \tau}_{1st \hskip 0.1cm mol.}  =  \frac{8 \epsilon_{0}
\epsilon_{1}^{\nu} \epsilon_{2}^{\mu}}{r^{2}}[-3 \sigma^{3} \lambda_{D}^{2} (1-\
\lambda_{D}) ({\bf k}.{\bf M} \times 
{\bf k})  \nonumber \\ 
(2 \rho^{-13} - \rho^{-7})/ 
\sigma_{0} 
+ \lambda_{E}^{2} (1-\lambda_{E}) 
({\bf k}_{E}.{\bf A} \times {\bf k}_{E}) \nonumber \\
(\rho^{-12} - \rho^{-6}) ]  
 + \nu U \epsilon: {\bf E}^{-1}{\bf M} \nonumber \\.
\end{eqnarray}
($\epsilon$ in equation (12) is the Levi-Civita tensor)

%\section{Results of simulation studies}

To check the suitability of the proposed model, we performed a sample NVE Molecular Dynamics (MD) simulation considering a system of prolate biaxial ellipsoids, interacting through the model potential described in the previous section. 
In this simulation we take two systems having different aspect ratios and well depth ratios.  The systems consisting of 1372 ellipsoids interacting through this new potential all have $\mu = 1, \nu = 2$. The other set of parameters are taken as follows:\\
The ratios $\sigma_{x}:\sigma_{y}:\sigma_{z}$ have values $1:1.5:4.5$ and $1:2:4.5$ % and $1:2.5:4.6$ 
and  corresponding $I^{*}_{xx},I^{*}_{yy},I^{*}_{zz}$ have the set of values $(1.125,1.062,0.16); (1.2125,1.062,0.25)
$ %; (1.325,1.062,0.36)$ 
for two systems respectively. All the systems have $\epsilon_{x}/\epsilon_{z} = 1/30$
but having different  $\epsilon_{y}/\epsilon_{z}$ values $1/7, 1/8.5$  for two systems respectively.

In the NVE MD simulation, 1372 molecules were confined in the cubic box obeying minimum image convention. The run started from a density
$\rho^{*} = \rho \sigma_{0}^{3}$ = 0.01 with ellipsoidal molecules
loacted on the sites of FCC lattice and having parallel
orientation. For MD initial step
translational and angular velocities were assigned
from the Gaussian distribution of velocities at a higher temperature. This structure was not a stable structure at this density
and it was melted at a reduced temperature $T^{*} = k_{B}T/
\epsilon_{0} = 4.0$ .
We used this isotropic
configuration which was both orientationally and translationally
disordered, as the initial configuration. Forces and torques were calculated
from equations (12) and (13) using leap-frog algorithm \cite{All_Tild}
.   
To get the most ordered structure for the system, we increased the density from
$\rho^{*} = 0.01$ to the desired values of $\rho^{*} = 0.19$ and $0.16$ 
%and 0.13$ 
respectively for two set of systems with an increament size
of $0.001$ upto $\rho^{*} = 0.1$ and $0.01$ for the rest. Temperature was
then lowered in steps to help the molecules set with more order.
The systems were cooled down to a very low reduced temperature $T^{*} = k_{B}T/
\epsilon_{0} = 0.4$.
%For each system this process required altogether $5 \times 10^{6}$ MC cycles for crystallization. 
The interaction potential was taken equal to half the simulation box length.  The time step size $\delta t^{*} =
\delta t/(m \sigma_{0}^{2} / \epsilon_{0})^{1/2} = 0.0012$  during cooling process. 
The orientations of molecules were described by quaternions
instead of Eulerian angles to get the singularity-free orientational
equations of motion. 

To investigate the phase  behaviour we started increasing the temperature
$T^{*}$  gradually. For each temperature we calculated
the orientational order parameters, correlation functions  to identify the particular liquid crystal phase. To achieve an equilibrated higher temperature 
configuration from more ordered low temperature phase $ 2 \times 10^{5}$ cycles 
were required far from transition and $ 5 \times 10^{5}$ near transition.

The orientational order parameter for uniaxial phase was calculated
from the largest eigen value obtained by diagonalization of the order
parameter tensor

\begin{equation}
Q_{\alpha \beta} = \frac{1}{2 N} \sum(3 e_{i \alpha} e_{i \beta}
- \delta_{\alpha \beta})\hskip 1.8cm  \alpha, \beta = x,y,z.
\end{equation}

where $e_{i \alpha}$ was the $\alpha$ th component of the unit vector
$e_{i}$ along the symmetry axis of the i th molecule. Corresponding
eigenvector gave the director which defines the average direction
of molecular alignment. 

The biaxial order parameter is $ \langle R_{2,2}^{2}\rangle=\langle\frac{1}{2}(1+\cos
^{2}\beta) \cos 2\alpha \cos 2\gamma -\cos\beta \sin2 \alpha \sin2 \gamma 
\rangle$.

Additionally, to detect phase behaviour more 
accurately, some structural quantities were calculated. Those were
 radial distribution function
$g(r) =  < \delta (r-r_{ij})>_{ij}/ 4 \pi r^{2} \rho$; density along
the director $g(z) =  < \delta (z-z_{ij})>_{ij} / \pi R^{2} \rho $,
where $z_{ij} = r_{ij} cos \beta_{r_{ij}}$ was measured in the
director frame and $R$ is the radius of the cylindrical sampling region.

The snapshots of the simulation study show various phases formed at different temperatures. At $T^{*}=0.4$ the smectic phase shows biaxiality in both cases as seen in the structures for the first system in Fig.1 and Fig.2 and for the second system in Fig.3 and Fig.4. At $T^{*}= 1.0$ it shows biaxial nematic phases in Fig.5 and Fig.6 for the first system and fig.7 and Fig.8  for the second system. At $T^{*}>5.0$ the systems becomes isotropic. The radial distribution functions $g(r)$ (Fig.9, Fig.11) and density projection to the director $g(z)$ (Fig.10, Fig.12) are plotted at different temperatures for both systems. The values of orientational order parameters $\eta$ and $R_{22}$ are plotted in Fig.13 and Fig.14. Our results show that the new model potential can successfully reproduce both  biaxial smectic and  biaxial nematic phases. To demonstrate this new model's ability to generate biaxial liquid crystal phases we are reporting here some preliminary results. We plan to communicate details of phase transitional behaviour of this system in future work.

%\section{Results}
%\label{sec:results}

%\section{Conclusion}
%\label{sec:conclusion}

Liquid crystals show a rich multiplicity of phases which comes from their 
complex molecular structures giving rise to complex electromagnetic interactions.
Other molecules of biological importance e.g. base pairs of DNA, lipids etc.
basically have anisotropic biaxial structures. It is a difficult task to
obtain the distance of closest approach as a function of orientation for these 
molecules. Additionally, this potential must capture the energetic strength anisotropy 
in a consistent way. GOP is a widely used model potential for the description of anisotropic behaviour of liquid crystals, however, it does not give a proper geometric
interpretation. It can calculate the proper range and strength parameters for some specific configurations of identical uniaxial molecules.
Perram and Wertheim took a different approach by introducing the ellipsoid contact function which took care of its true geometrical aspect by doing correct estimation of position of contact. 
Although applicable to any mixture in geometrical context it did not distinguish in energy functions between different relative orientations and positions of the non-spherical molecular pair.
As this energy function is the basis of well depth estimation, it provides 
necessary and sufficient weightage while determining proper magnitude of the attaractive part of the pair potential and thus  acts as the most important contributor in fine tuninng of the potential so that it generates a specific phase structure. In this
work, considering molecular orientations and positions explicitly in both structural and energy strength related terms, we modelled
a generalized pair potential for biaxial molecules. We further studied liquid crystal
phases by doing molecular dymanics simulation utilizing this
model potential. Though biaxial phases have been synthesized and analysed recently, theoretical studies including simulations have been done for more than three decades to understand the molecular properties needed for stabilization of these amazing phases. However, underlying intermolecular interactions responsible for the generation of these phase are still not understood properly and the present model potential will be able to provide essential informations about the characteristic of single component rod-like molecular system crucial for obtaining biaxial phases. The simulation results show that this new model potential reported in this paper is qualitatively suitable to reproduce complex mesogenic behaviour of organic biaxial molecules in a realistic way. This new  potential has been generalized to study systems comprising of multicomponent dissimilar molecules, results of which will be communicated in future publication.

%\begin{acknowledgments}

%T.K. Bose gratefully acknowledges the support of CSIR, India for providing Jun
%ior Research Fellowship via sanction no. 09/028(0794)/2010-EMR-I. 
The graphics
software QMGA \cite{QMGA} was used for producing snapshots of the phases. 
This work 
was supported by UGC-UPE scheme of the University of Calcutta.

%\end{acknowledgments}
%\nocite{*}
\bibliographystyle{plainnat}.

\bibliography{PRE}

\begin{figure}
\begin{center}
%\centering
\includegraphics[angle=0.0,scale=0.7]{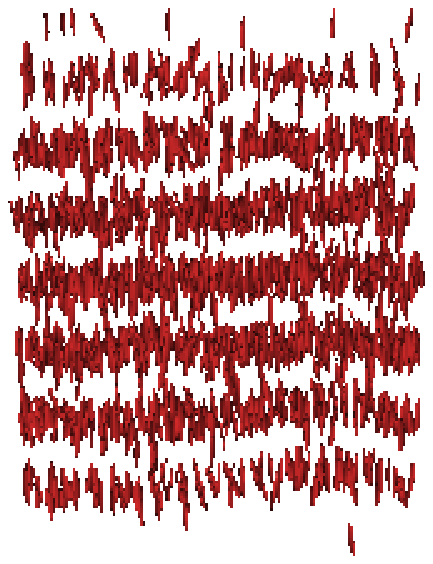}
%\epsfxsize=4.0in
%\epsfysize=3.8in
%\rotatebox{0}{\epsfbox{fig1.eps}}
\end{center}
\caption{Snapshot at $T^{*}=0.4$  for 
$\sigma_{x}:\sigma_{y}:\sigma_{z} = 1:1.5:4.5$}
\label{figure.1}
\end{figure}

% ------------------------------------
\begin{figure}
\begin{center}
%\centering
\includegraphics[angle=0.0,scale=1.0]{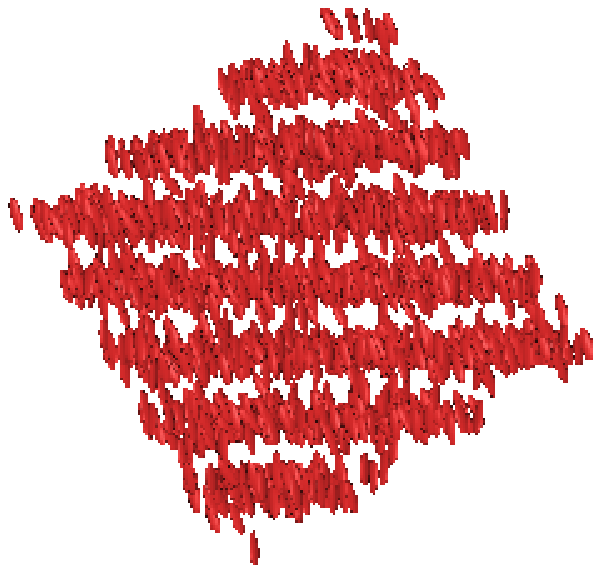}
%\epsfxsize=4.0in
%\epsfysize=3.8in
%\rotatebox{0}{\epsfbox{fig1.eps}}
\end{center}
\caption{Snapshot (rotated) at $T^{*}=0.4$  for 
$\sigma_{x}:\sigma_{y}:\sigma_{z} = 1:1.5:4.5$}
\label{figure.2}
\end{figure}

% ------------------------------------
\newpage
\begin{figure}
\begin{center}
%\centering
\includegraphics[angle=0.0,scale=1.0]{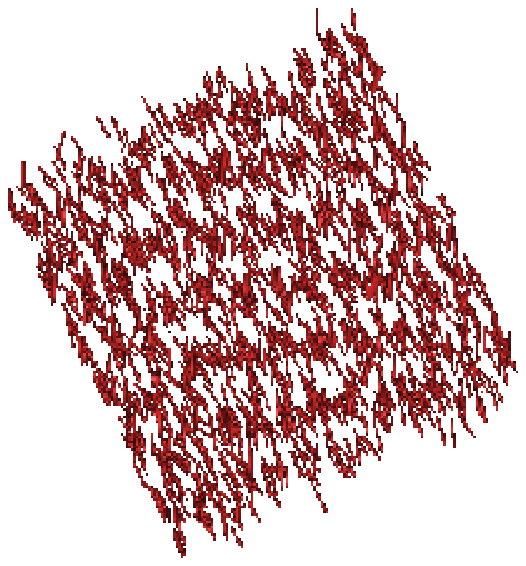}
%\epsfxsize=4.0in
%\epsfysize=3.8in
%\rotatebox{0}{\epsfbox{fig1.eps}}
\end{center}
\caption{Snapshot at $T^{*}=0.4$  for 
$\sigma_{x}:\sigma_{y}:\sigma_{z} = 1:2.0:4.5$}
\label{figure.3}
\end{figure}
%\vspace*{3.0cm}
% ------------------------------------
\begin{figure}
\begin{center}
%\centering
\includegraphics[angle=0.0,scale=0.9]{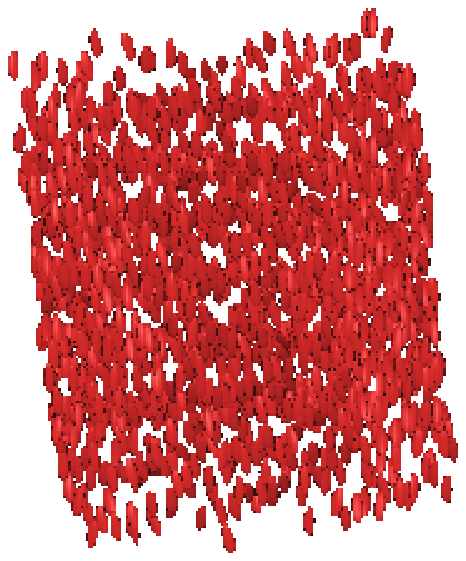}
%\epsfxsize=4.0in
%\epsfysize=3.8in
%\rotatebox{0}{\epsfbox{fig1.eps}}
\end{center}
\caption{Snapshot (rotated) at $T^{*}=0.4$  for 
$\sigma_{x}:\sigma_{y}:\sigma_{z} = 1:2.0:4.5$}
\label{figure.4}
\end{figure}

% ------------------------------------
\newpage
\begin{figure}
\begin{center}
%%\centering
\includegraphics[angle=0.0,scale=1.0]{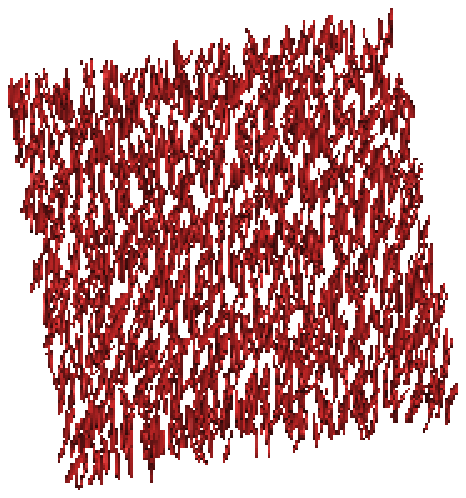}
%%\epsfxsize=4.0in
%%\epsfysize=3.8in
%%\rotatebox{0}{\epsfbox{fig1.eps}}
\end{center}
\caption{Snapshot at $T^{*}=1.0$  for 
$\sigma_{x}:\sigma_{y}:\sigma_{z} = 1:1.5:4.5$}
\label{figure.5}
\end{figure}

% ------------------------------------
\begin{figure}
\begin{center}
%\centering
\includegraphics[angle=0.0,scale=1.0]{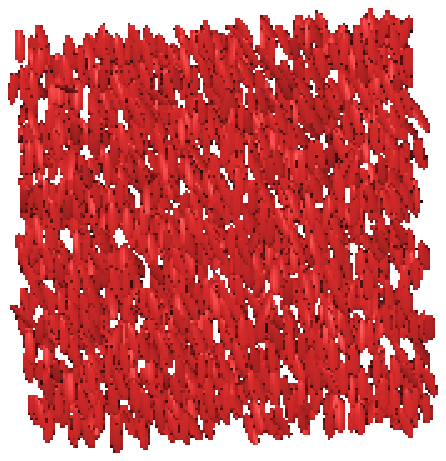}
%\epsfxsize=4.0in
%\epsfysize=3.8in
%\rotatebox{0}{\epsfbox{fig1.eps}}
\end{center}
\caption{Snapshot (rotated) at $T^{*}=1.0$  for 
$\sigma_{x}:\sigma_{y}:\sigma_{z} = 1:1.5:4.5$}
\label{figure.6}
\end{figure}
% ------------------------------------
\vspace*{5.0cm}
\newpage
\begin{figure}
\begin{center}
%\centering
\includegraphics[angle=0.0,scale=1.0]{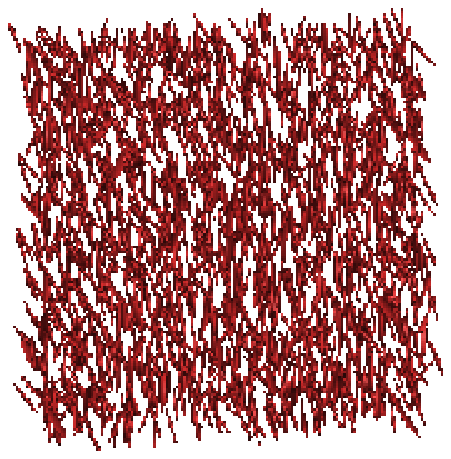}
%\epsfxsize=4.0in
%\epsfysize=3.8in
%\rotatebox{0}{\epsfbox{fig1.eps}}
\end{center}
\caption{Snapshot (rotated) at $T^{*}=1.0$  for 
$\sigma_{x}:\sigma_{y}:\sigma_{z} = 1:2.0:4.5$}
\label{figure.7}
\end{figure}

% ------------------------------------
\begin{figure}
\begin{center}
%\centering
\includegraphics[angle=0.0,scale=1.0]{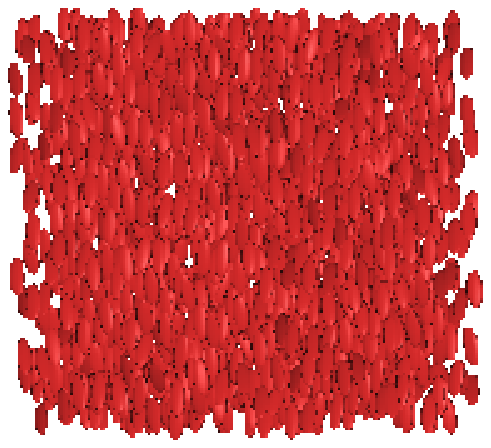}
%\epsfxsize=4.0in
%\epsfysize=3.8in
%\rotatebox{0}{\epsfbox{fig1.eps}}
\end{center}
\caption{Snapshot (rotated) at $T^{*}=1.0$  for 
$\sigma_{x}:\sigma_{y}:\sigma_{z} = 1:2.0:4.5$}
\label{figure.8}
\end{figure}

% ------------------------------------
\newpage
\begin{figure}
\begin{center}
%\centering
\includegraphics[angle=270.0,scale=0.7]{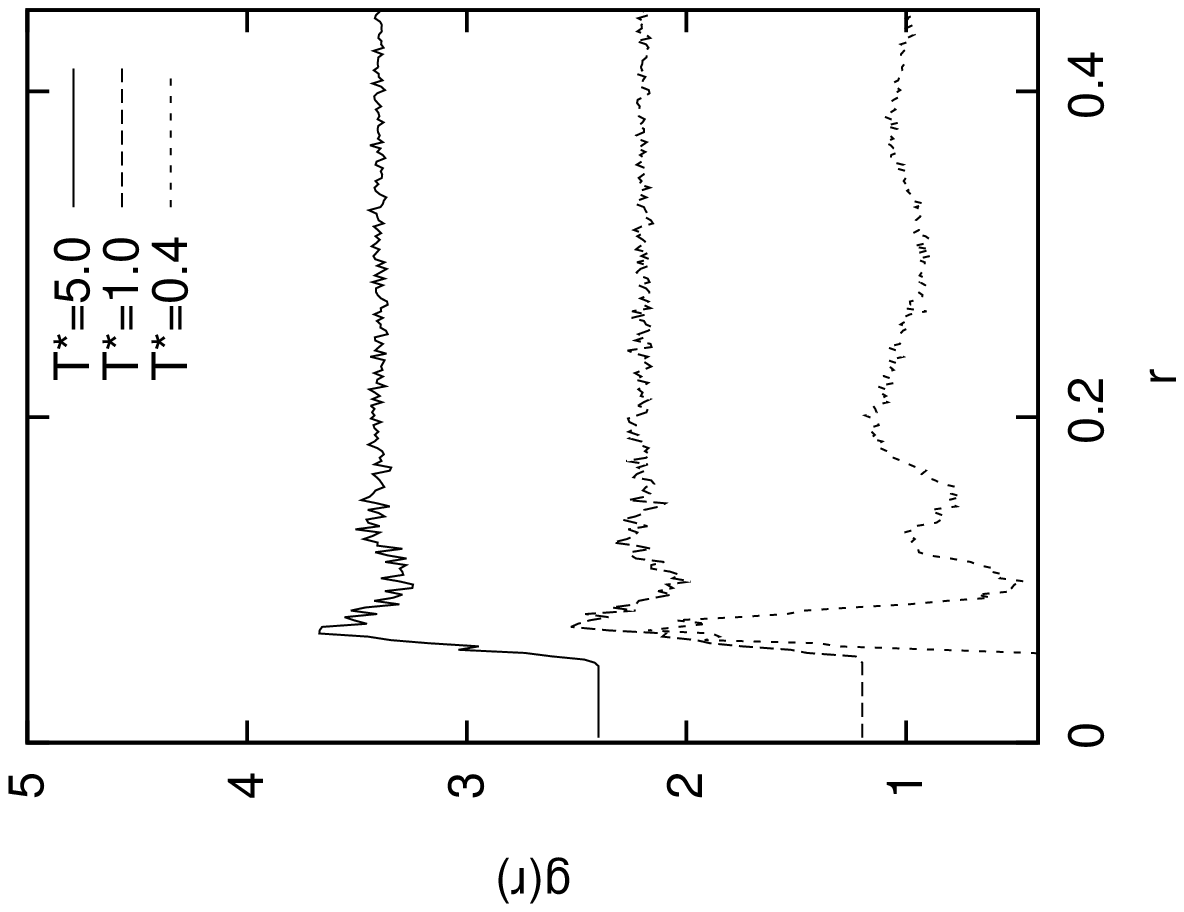}
%\epsfxsize=4.0in
%\epsfysize=3.8in
%\rotatebox{0}{\epsfbox{fig1.eps}}
\end{center}
\caption{Radial distribution function g(r) for 
$\sigma_{x}:\sigma_{y}:\sigma_{z} = 1:1.5:4.5$}
\label{figure.9}
\end{figure}

% ---------------------------------------------------
\begin{figure}
\begin{center}
%\centering
\includegraphics[angle=270.0,scale=0.7]{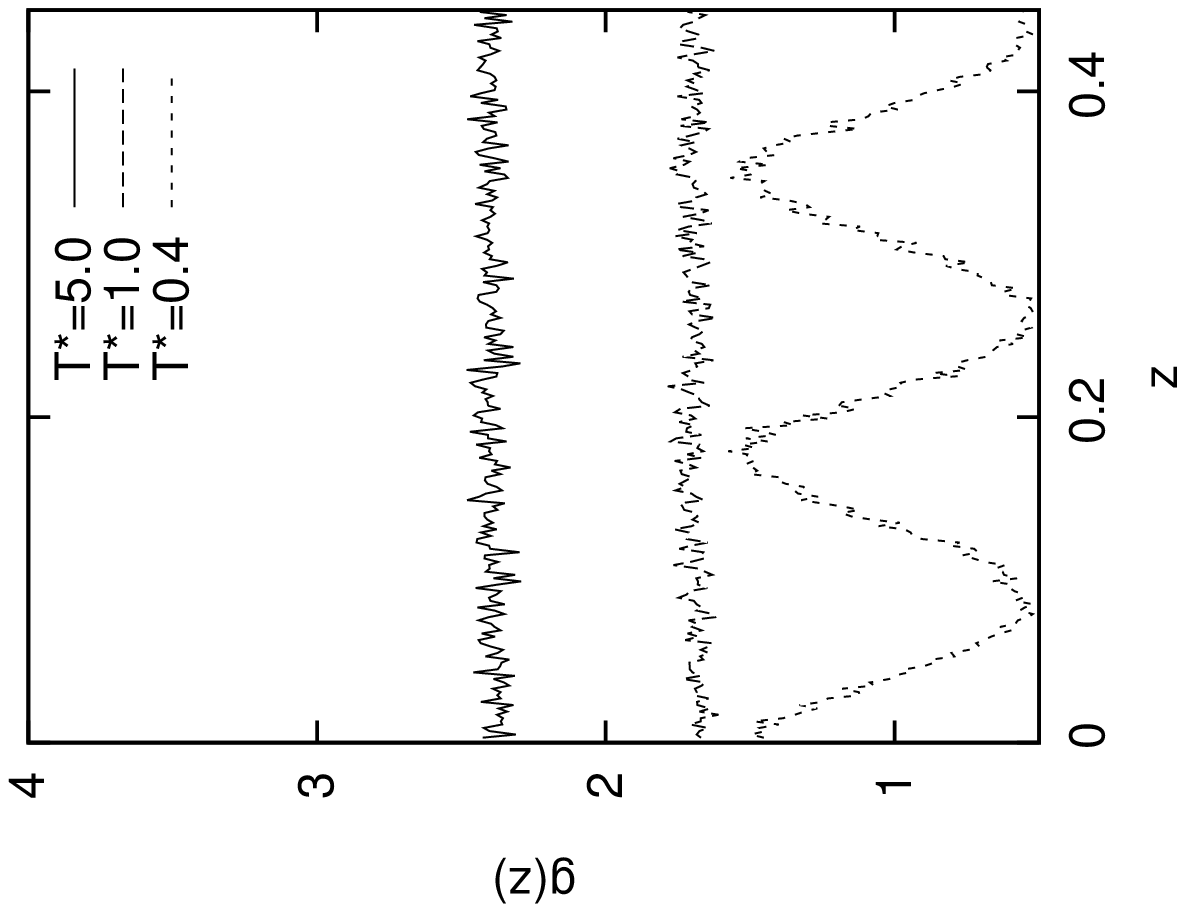}
%\epsfxsize=4.0in
%\epsfysize=3.8in
%\rotatebox{0}{\epsfbox{fig1.eps}}
\end{center}
\caption
{Denity projection with respect to the director g(z) for $\sigma_{x}:\sigma_{y}:\sigma_{z} = 1:1.5:4.5$}
\label{figure.10}
\end{figure}

% ------------------------------------------
\newpage
\begin{figure}
\begin{center}
%\centering
\includegraphics[angle=270.0,scale=0.7]{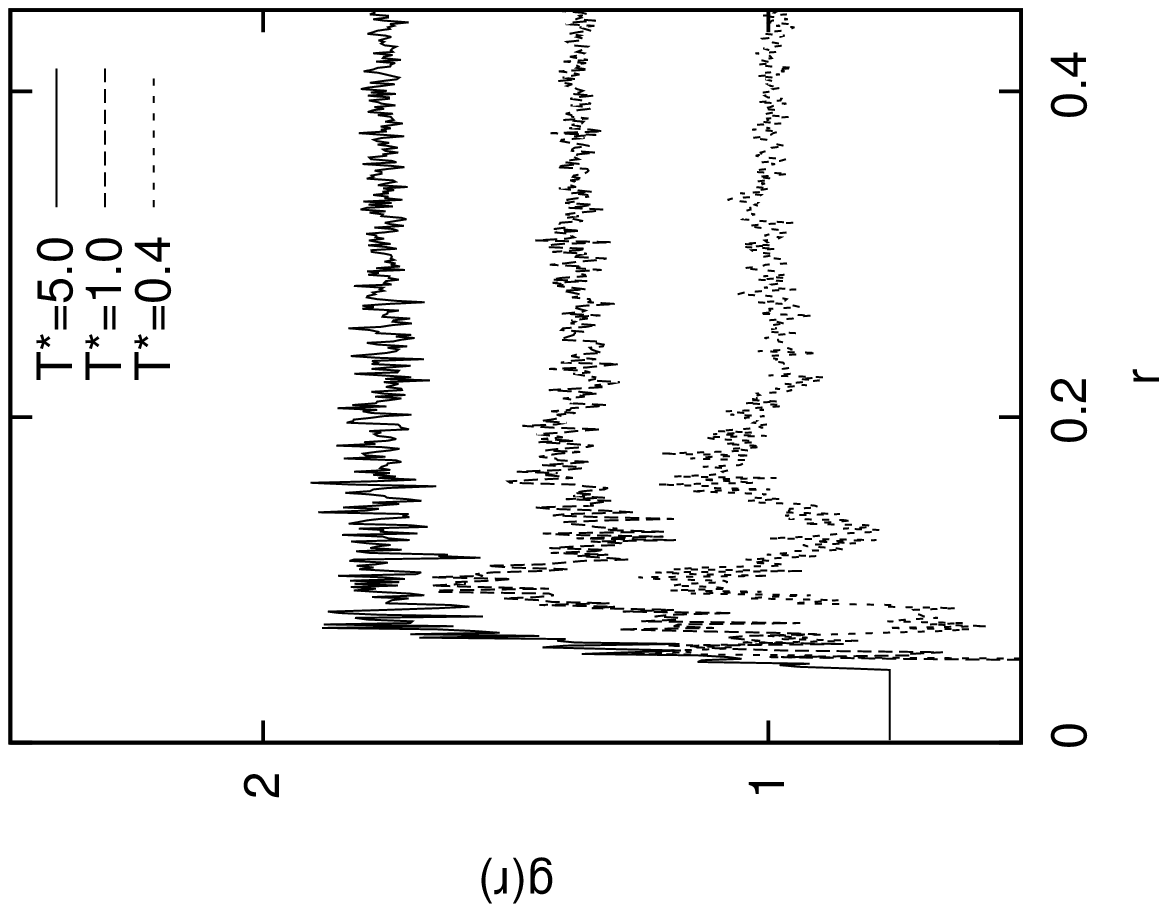}
%\epsfxsize=4.0in
%\epsfysize=3.8in
%\rotatebox{0}{\epsfbox{fig1.eps}}
\end{center}
\caption{Radial distribution function g(r) for $\sigma_{x}:\sigma_{y}:\sigma_{z} = 1:2:4.5$ }
\label{figure.11}
\end{figure}

% ---------------------------------------------------
\begin{figure}
\begin{center}
%\centering
\includegraphics[angle=270.0,scale=0.7]{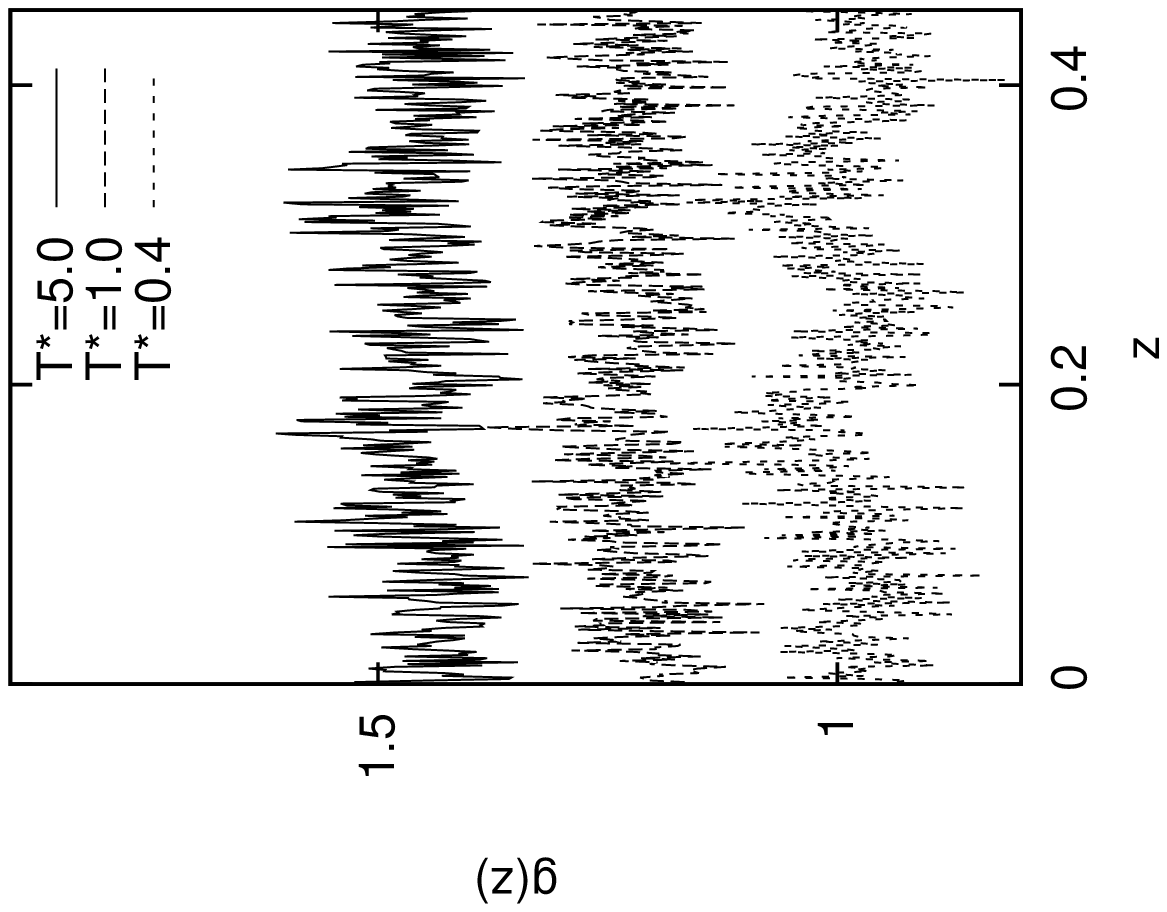}
%\epsfxsize=4.0in
%\epsfysize=3.8in
%\rotatebox{0}{\epsfbox{fig1.eps}}
\end{center}
\caption
{Denity projection with respect to the director g(z) for $\sigma_{x}:\sigma_{y}:\sigma_{z} = 1:2:4.5$}
\label{figure.12}
\end{figure}

%-----------------------------------------------------
\newpage
\begin{figure}
\begin{center}
%\centering
\includegraphics[angle=270.0,scale=0.7]{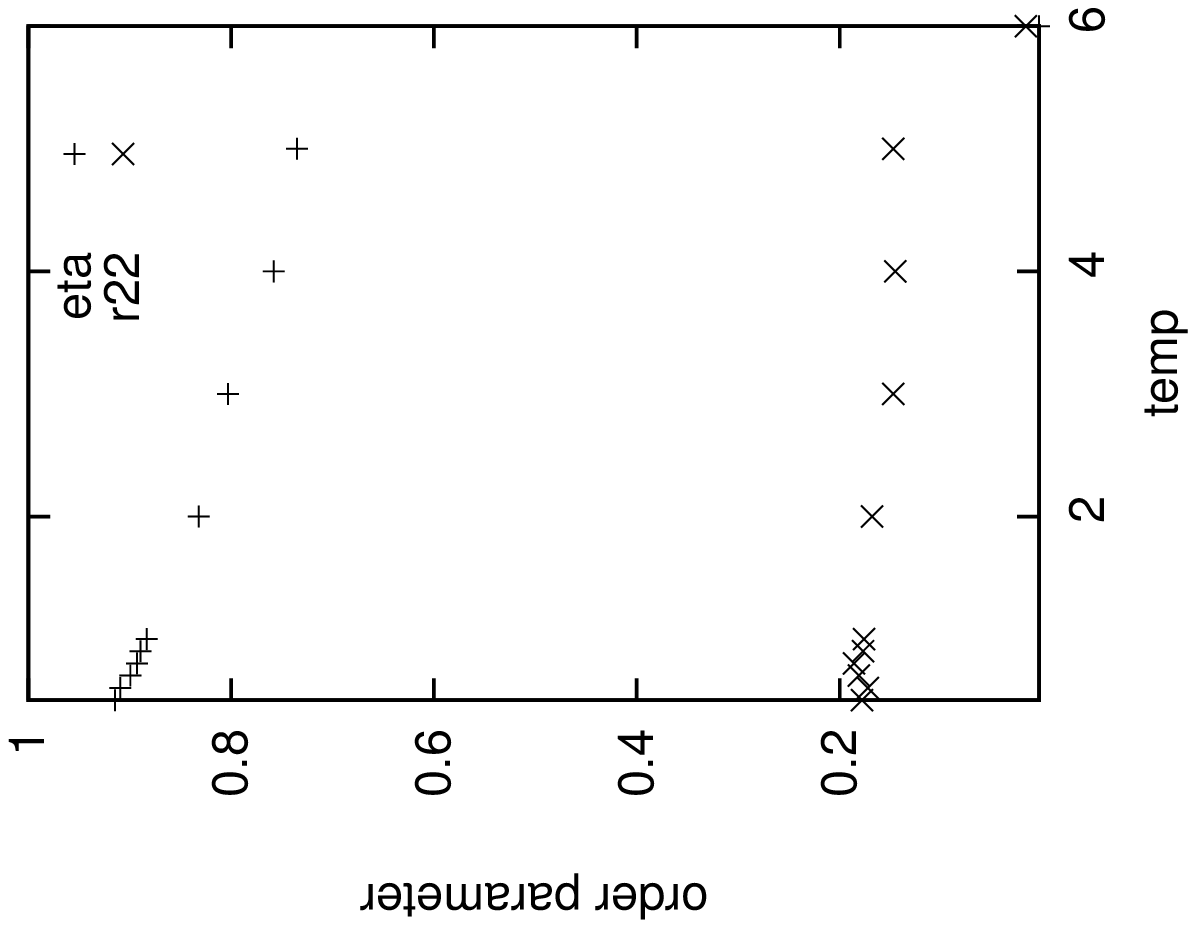}
%\epsfxsize=4.0in
%\epsfysize=3.8in
%\rotatebox{0}{\epsfbox{fig1.eps}}
\end{center}
\caption
{order parameters for $\sigma_{x}:\sigma_{y}:\sigma_{z} = 1:1.5:4.5$}
\label{figure.13}
\end{figure}

%-----------------------------------------------------
\begin{figure}
\begin{center}
%\centering
\includegraphics[angle=270.0,scale=0.7]{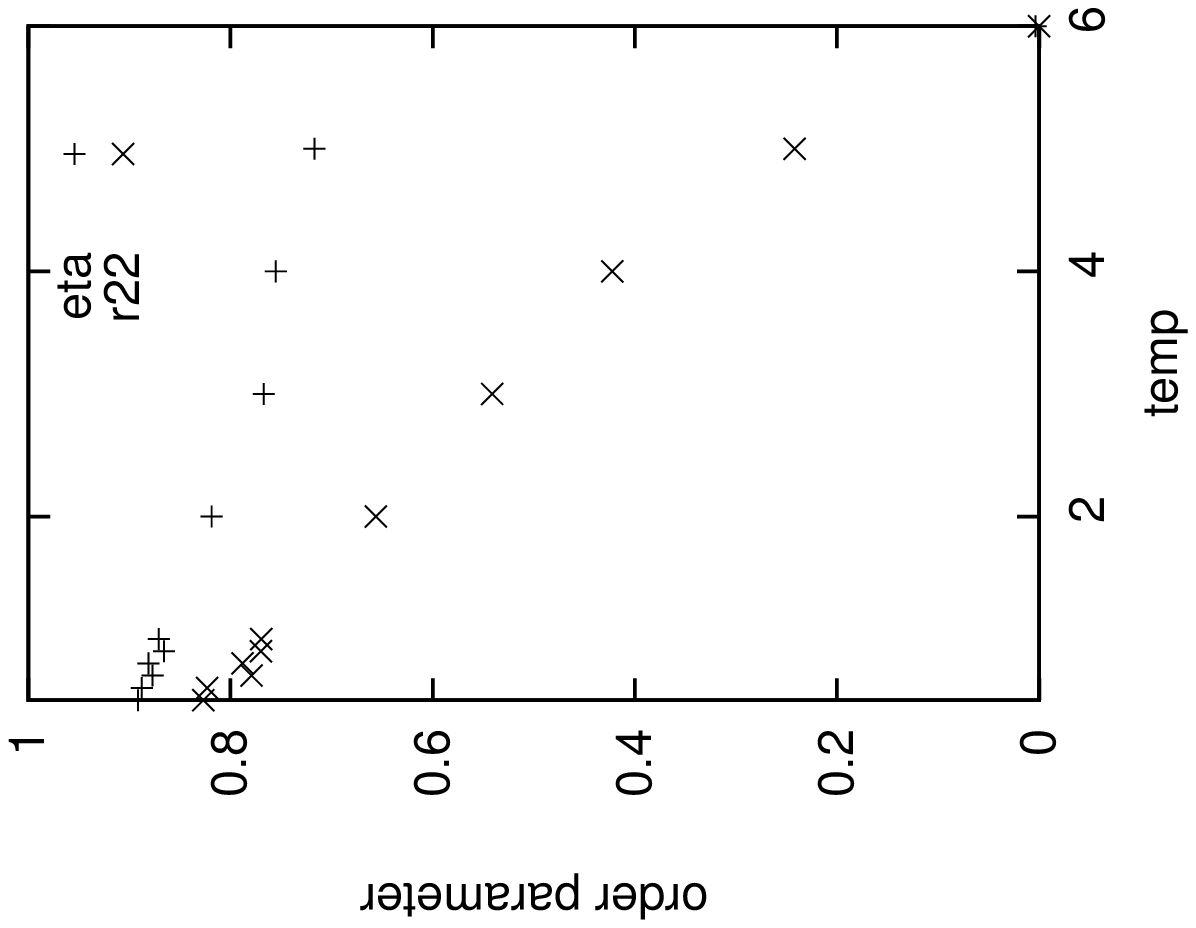}
%\epsfxsize=4.0in
%\epsfysize=3.8in
%\rotatebox{0}{\epsfbox{fig1.eps}}
\end{center}
\caption
{order parameters for $\sigma_{x}:\sigma_{y}:\sigma_{z} = 1:2:4.5$}
\label{figure.14}
\end{figure}
\end{document}